\documentclass[a4paper,11pt]{article}
\usepackage{pos}
\usepackage{bbm,color}
\usepackage{mathtools}
\usepackage{hyperref}
\usepackage{graphicx}
\usepackage{epsfig}
\usepackage{euscript}
\usepackage{xcolor}
\usepackage{fancybox}
\usepackage{enumerate}
\usepackage{widetext}
\usepackage{subfigure}
\definecolor{brightturquoise}{rgb}{0.03, 0.91, 0.87}
\definecolor{awesome}{rgb}{1.0, 0.13, 0.32}
\definecolor{armygreen}{rgb}{0.29, 0.33, 0.13}
\definecolor{aqua}{rgb}{0.0, 1.0,1.0}
\definecolor{maroon(html/css)}{rgb}{0.5, 0.0,0.0}
\definecolor{pinegreen}{rgb}{0.0, 0.47,0.44}
\definecolor{red-brown}{rgb}{0.65, 0.16,0.16}

\newcommand{\bhlumi}{{\tt BHLUMI}}
\newcommand{\bhwide}{{\tt BHWIDE}}
\newcommand{\babayaga}{{\tt BabaYaga}}
\newcommand{\order}[1]{${\cal O}(#1)$}

\newcommand{\sfac}{\mathfrak{s}}

\title{Overview of the Path to 0.01$\%$ Theoretical Luminosity
Precision for the FCCee and Its Possible Synergistic Effects
for Other FCC Precision Theory Requirements}
\ShortTitle{Overview: Path to 0.01$\%$ Theoretical Luminosity
Precision}

\author*[a]{B.F.L. Ward}
\author[b]{S. Jadach}
\author[c]{W. Placzek}
\author[b]{M. Skrzypek}
\author[d]{S. A. Yost}
\affiliation[a]{Baylor University,\\
  Waco, TX, USA}

\affiliation[b]{Institute of Nuclear Physics Polish Academy of Sciences,\\
Krakow, PL}

\affiliation[c]{Institute of Applied Computer Science, Jagiellonian University,\\
Krakow, PL}

\affiliation[d]{The Citadel,\\
Charleston, SC, USA}

\emailAdd{bfl\_ward@baylor.edu}
\emailAdd{Stanislaw.Jadach@cern.ch}
\emailAdd{Wieslaw.Placzek@uj.edu.pl}
\emailAdd{Maciej.Skrzypek@cern.ch}
\emailAdd{yosts1@citadel.edu}        
                                                   
\abstract{To exploit properly the precision physics program at the FCC-ee, the theoretical precision tag on the respective luminosity will need to be improved from the 0.054$\%$ (0.061$\%$) results at LEP to 0.01$\%$, where the former (latter) LEP result has (does not have) the pairs correction. We present an overview of the roads one may take to reach the required 0.01$\%$ precision tag at the FCC-ee and we discuss possible synergistic effects of the walk along these roads for other FCC precision theory requirements}
\centerline{BU-HEPP-20-07, Dec., 2020}
\FullConference{%
  40th International Conference on High Energy physics - ICHEP2020\\
  July 28 - August 6, 2020\\
  Prague, Czech Republic (virtual meeting)
}

\begin{document}
\maketitle

\baselineskip=11pt
\section{Introduction}
It is well documented~\cite{fccwksp2019,fccwksp2018,blondel-jnt-2019} by now the new era of precision for Higgs and EW observables which will obtain at the proposed precision physics factories such as the FCC-ee, CLIC, ILC and CEPC. These scenarios all require that the theoretical luminosity precision be improved from its LEP era value to the regime of 0.01\%, or better if possible\footnote{In addition to this improvement, these new precision scenarios may require a completely new methodology of the
QED “deconvolution” and its related new definition
of the EW pseudo-observables (EWPO’s)~\cite{jad-skrpk-2019,gluza-2020}}. Accordingly,
the  general context of our discussion can be set by recalling the situation that existed at the end of LEP. At that time, as we show here in Table ~\ref{tab:error99}, the error budget for the \bhlumi4.04 MC used by all LEP collaborations to simulate the luminosity process was calculated in Ref.~\cite{bhlumi-precision:1998}. 
\begin{table}[!ht]
\centering
\begin{tabular}{|l|l|l|l|l|l|}
\hline 
    & \multicolumn{2}{|c|}{LEP1} 
              & \multicolumn{2}{|c|}{LEP2}
\\ \hline 
Type of correction/error
    & 1996
         & 1999
              & 1996
                   & 1999
\\  \hline 
(a) Missing photonic ${\cal O}(\alpha^2 )$~\cite{Jadach:1995hy,Jadach:1999pf} 
    & 0.10\%      
        & 0.027\%    
            & 0.20\%  
                & 0.04\%
\\ 
(b) Missing photonic ${\cal O}(\alpha^3 L_e^3)$~\cite{Jadach:1996ir} 
    & 0.015\%     
        & 0.015\%    
            & 0.03\%  
                & 0.03\% 
\\ 
(c) Vacuum polarization~\cite{BW9,BW10} 
    & 0.04\%      
        & 0.04\%    
           & 0.10\%  
                & 0.10\% 
\\ 
(d) Light pairs~\cite{BW11,BW12} 
    & 0.03\%      
        & 0.03\%    
            & 0.05\%  
                & 0.05\% 
\\ 
(e) $Z$ and $s$-channel $\gamma$~\cite{BW13,BW6}
    & 0.015\%      
        & 0.015\%   
            &  0.0\%  
                & 0.0\% 
\\ \hline 
Total  
    & 0.11\%~\cite{BW6}
        & 0.061\%~\cite{bhlumi-precision:1998}
            & 0.25\%~\cite{BW6}
                & 0.12\%~\cite{bhlumi-precision:1998}
\\ \hline 
\end{tabular}
\caption{\sf
Summary of the total (physical+technical) theoretical uncertainty
for a typical calorimetric detector.
For LEP1, the above estimate is valid for a generic angular range
within   $1^{\circ}$--$3^{\circ}$ ($18$--$52$ mrads), and
for  LEP2 energies up to $176$~GeV and an
angular range within $3^{\circ}$--$6^{\circ}$.
Total uncertainty is taken in quadrature.
Technical precision included in (a).
}
\label{tab:error99}
\end{table}
In this table, we cite the published works upon which the various error estimates are based as they are discussed in Ref.~\cite{bhlumi-precision:1998}. From what we show in this table, we can chart a path to 0.01\% luminosity theory error. In discussing this path, we will use the FCC-ee as our prototype but it will be clear that the arguments apply as well to the other similar precision EW factories.\par
As an example, consider the role of the Bhabha luminosity(${\cal L}$) measurement in neutrino counting where we use $N_\nu$ to denote the number of light neutrinos. From the LEP Z-peak measurements~\cite{adlo-sld1}, we have 
\begin{equation}
\begin{split}
N_\nu&=2.9840\pm 0.0082 \nonumber\\
\delta N_\nu &= 10.5\frac{\delta n_{had}}{n_{had}}\oplus3.0\frac{\delta n_{lept}}{n_{lept}}\oplus7.5\frac{\delta{\cal L}}{\cal L}\nonumber\\
\frac{\delta{\cal L}}{\cal L}&=0.61\%\Rightarrow \delta N_\nu = 0.0046.
\end{split}
\end{equation}
Recently, Janot and Jadach~\cite{Janot:2019oyi}, building on the results in Ref.~\cite{Voutsinas:2019hwu}, use the current status of the LEP luminosity theory error together with improved measurement error analysis correcting for so-far ignored systematic beam-beam effects to show that $N_\nu=2.9963\pm 0.0074$ with $\delta N_\nu = 0.0028$ from $\frac{\delta{\cal L}}{\cal L}$.
This resolves what had been a lingering ``2$\sigma$'' deviation from the predictions of the Standard Theory~\cite{djg-smat50}. 
\par
\section{Path to 0.01\% Luminosity Theory Uncertainty}
The 0.01\% precision tag needed for the luminosity theory error for the FCC-ee can be realized by developing the corresponding improved version
of \bhlumi. This problem is discussed in some detail in Ref.~\cite{Jadach:2018}, wherein the path to 0.01\% theory precision for the FCC-ee luminosity is 
presented. In Table ~\ref{tab:lep2fcc}, the results of this latter reference are shown, wherein we also present the current state of the art for completeness, as it is discussed in more detail in Refs.~\cite{Jadach:2018}.
\begin{table}[ht!]
\centering
\begin{tabular}{|l|l|l|l|}
\hline
Type of correction~/~Error
        & Update 2018
                &  FCC-ee forecast
\\ \hline 
(a) Photonic $[{\cal O}\left(L_e\alpha^2 \right)]\; {\cal O}\left(L_e^2\alpha^3\right)$
        & 0.027\%
                &  $ 0.1 \times 10^{-4} $
\\ 
(b) Photonic $[{\cal O}\left(L_e^3\alpha^3\right)]\; {\cal O}\left(L_e^4\alpha^4\right)$
        & 0.015\%
                & $ 0.6 \times 10^{-5} $
\\
(c) Vacuum polariz.
        & 0.014\%~\cite{JegerlehnerCERN:2016}
                & $ 0.6 \times 10^{-4} $
\\
(d) Light pairs
        & 0.010\%~\cite{ON1,ON2}
                & $ 0.5 \times 10^{-4} $
\\
(e) $Z$ and $s$-channel $\gamma$ exchange
        & 0.090\%~\cite{BW13}
                & $ 0.1 \times 10^{-4} $
\\ 
(f) Up-down interference
    &0.009\%~\cite{Jadach:1990zf}
        & $ 0.1 \times 10^{-4} $
\\
(f) Technical Precision & (0.027)\% 
                & $ 0.1 \times 10^{-4} $
\\ \hline 
Total
        & 0.097\%
                & $ 1.0 \times 10^{-4} $
\\ \hline 
\end{tabular}
\caption{\sf
Anticipated total (physical+technical) theoretical uncertainty 
for a FCC-ee luminosity calorimetric detector with
the angular range being $64$--$86\,$mrad (narrow), near the $Z$ peak.
Description of photonic corrections in square brackets is related to 
the 2nd column.
The total error is summed in quadrature.
}
\label{tab:lep2fcc}
\end{table}
In arriving at Table~\ref{tab:lep2fcc} the main points are as follows. The errors of the photonic corrections in lines (a) and (b) in the LEP results in Table~\ref{tab:error99} are due to known effects from Refs.~\cite{Jadach:1995hy,Jadach:1999pf,Jadach:1996ir} that were not implemented into \bhlumi. We show in Table~\ref{tab:lep2fcc} what these errors will become after these known results are included in \bhlumi\ as discussed in Ref.~\cite{Jadach:2018}. Similarly, in line (c) of Table~\ref{tab:error99} the error is due to the LEP era uncertainty on the hadronic contribution to the vacuum polarization for the photon at the respective momentum transfers for the luminosity process; in Table~\ref{tab:lep2fcc} we show the expected improvement of this error for the FCC-ee as discussed in Refs.~\cite{JegerlehnerCERN:2016,Jegerlehner:2017zsb,jegerlhnr-Fccee-2019}.
\par
In line (d) in Table~\ref{tab:lep2fcc}, proceeding in a similar way, we show the expected~\cite{Jadach:2018} improvement, relative to the LEP time for Table~\ref{tab:error99}, in the light pairs error for the FCC-ee . As we explain in Ref.~\cite{Jadach:2018}, because non-photonic graphs can contribute as much as $0.01\%$ for the cut-off $z_{\rm cut} \sim 0.7$, the complete matrix element for the additional real $ e^{+} e^{-} $ pair radiation should be
used. This can be done with the MC generators developed for the $e^+ e^-\rightarrow 4f$ processes for  LEP2 physics - see Ref.~\cite{Jadach:2018} for further discussion. The contributions of light quark pairs, muon pairs and non-leading, non-soft additional $e^+e^- +n\gamma$ corrections can be controlled, with known methods~\cite{Jadach:2018}, such that the error on the pairs contribution is as given in line (d) for the FCC-ee. 
\par
We turn next to line (e) in Table~\ref{tab:lep2fcc}, wherein we show the improvement of the error on the $Z$ and $s$-channel $\gamma$ exchange for the FCC-ee as well as its current state of the art.
In Ref.~\cite{Jadach:2018}, a detailed discussion is presented of all of the six interference and three additional squared modulus terms  that result from the $s$-channel $\gamma$,  $s$-channel $Z$, and $t$-channel $Z$ exchange contributions to the amplitude for the luminosity process. It is shown that, if the predictions of \bhlumi\ for the luminosity
measurement at FCC-ee are combined with the ones from \bhwide\ ~\cite{bhwide:1997} for
this $Z$ and $s$-channel $\gamma$ exchange contribution, then the error in the second column of  line (e) of
Table~\ref{tab:lep2fcc} could be reduced to $0.01\%$. In order to reduce the uncertainty of this contribution
practically to zero we would include these $Z$ and $\gamma_s$ exchanges
within the CEEX~\cite{Jadach:2000ir} type matrix element at \order{\alpha^1} in \bhlumi. Here, CEEX stands for coherent exclusive exponentiation which acts at the level of the amplitudes
as compared the original Yennie--Frautschi--Suura~\cite{yfs:1961}(YFS)  exclusive exponentiation (EEX) that is used in \bhlumi4.04 and that acts at the level of the squared amplitudes.
It is expected to be enough to add the EW corrections
to the LABH process in the form of effective couplings in the Born amplitudes. This leads to the error estimate shown in Table~\ref{tab:lep2fcc} in line(e) for the FCC-ee.
\par
For completeness,  we note that for our discussion of the {\em Z} and {\em s}-channel $\gamma$ exchanges we made in Ref.~\cite{Jadach:2018} a numerical study using \bhwide\
for the the calorimetric LCAL-type
detector, as described in ref.~\cite{Jadach:1991cg},
for the symmetric angular range $64$--$86\,$mrad without
any cut on acoplanarity. The pure weak corrections were calculated with the  {\tt ALIBABA} 
EW library~\cite{Beenakker:1990mb,Beenakker:1990es}. The results, shown in Table ~\ref{tab:Zsgam},
were obtained for three values of the centre-of-mass (CM)
energy: $E_{\rm CM} = M_Z,\, M_Z\pm 1\,$GeV, where the latter two values have $Z$ contributions that are close to maximal in size.
\begin{table}[!ht]
\centering
\begin{tabular}{|c|c|c|c|c|}
\hline 
$E_{\rm CM}$ [GeV]
    & $\Delta_{\rm tot}$ [\%]
         & $\delta_{{\cal O}(\alpha)}^{\rm QED}$ [\%]
              & $\delta_{\rm h.o.}^{\rm QED}$ [\%]
                   & $\delta_{\rm tot}^{\rm weak}$ [\%]
\\  \hline 
$90.1876$       
  & $+0.642\,(12)$ 
    & $-0.152\,(59)$
      & $+0.034\,(38)$
        & $-0.005\,(12)$
\\ 
$91.1876$       
  & $+0.041\,(11)$ 
    & $+0.148\,(59)$
      & $-0.035\,(38)$
        & $+0.009\,(12)$
\\ 
$92.1876$       
  & $-0.719\,(13)$ 
    & $+0.348\,(59)$
      & $-0.081\,(38)$
        & $+0.039\,(13)$
\\ \hline 
\end{tabular}
\caption{\sf
Results from \bhwide\ for the $Z$ and $\gamma_s$ exchanges
contribution to the FCC-ee luminosity with respect to the
$\gamma_t\otimes\gamma_t$ process for the calorimetric
LCAL-type detector \cite{Jadach:1991cg} 
with the symmetric angular range $64$--$86\,$mrad; 
no acoplanarity cut was applied.
MC errors are marked in brackets.
}
\label{tab:Zsgam}
\end{table}
The results in the second column for the total size of the  $Z$ and $\gamma_s$ exchanges are consistent with our expectations as explained in ref.~\cite{Jadach:2018}: the contribution is positive below the $Z$ peak where it reaches a size $\sim 0.64\%$, is close to zero near the peak, and changes sign above the peak where it reaches a size $\sim -0.72\%$.
The third column, which features the fixed-order (non-exponentiated) ${\cal O}(\alpha)$ QED correction, shows that it is sizable and up to a half of the size of the Born level effect, but
with the opposite sign. The fourth column shows the size of the higher-order QED effects from YFS exponentiation, which also change their sign near the $Z$-peak, oppositely to the corresponding change of the ${\cal O}(\alpha)$ corrections. The size of the former effects is about
a quarter of that of the latter. From the effects in the fourth column we make a conservative estimate of the size of the missing higher-order QED effects in \bhwide\ using the big log factor $\gamma=\frac{\alpha}{\pi}\ln\frac{|\bar{t}|}{m_e^2}=0.042$ of Section 4 of Ref.~\cite{Jadach:2018} and a safety factor of 2 of Ref.~\cite{BW13} together with the largest higher-order effect  in Table~\ref{tab:Zsgam}, $0.081\%$, as $0.081\%\times \gamma \times 2 \simeq 0.007\%$. The last column shows that the size of the pure weak corrections is at the level of $0.01\%$ below and at $M_Z$ and increases up to $\sim 0.04\%$ above $M_Z$. We use the same factor as we did for the higher order corrections to estimate the size of the missing higher order pure weak corrections in \bhwide\ as $\sim 0.003\%$. Altogether, by adding the two estimates of its massing effects, we obtain a conservative estimate of $0.01\%$ for the physical precision of \bhwide\ to justify our earlier remarks concerning the error in line (e) of Table~\ref{tab:lep2fcc} that would result from the combination of the prediction of \bhlumi\ and that of \bhwide\ for this contribution. 
\par
In line(f) in Table~\ref{tab:lep2fcc} we show the estimate of the error on the up-down interference between radiation from the $e^-$ and $e^+$  lines. While it was negligible in LEP1, for the FCC-ee this effect, calculated in Ref.~\cite{Jadach:1990zf} at ${\cal O}(\alpha^1)$, is 10 times larger and has to be included in the upgraded \bhlumi. Once this is done, the error estimate shown in line(f) for the FCC-ee obtains~\cite{Jadach:2018}.
\par
This brings us to the issue of the technical precision. In order to get the upgraded \bhlumi\ technical precision
at the level $10^{-5}$ for the total cross section and $10^{-4}$
for single differential distributions, in an ideal situation, one would need to compare
it with another MC program developed independently,
which properly implements the soft-photon resummation,
LO corrections up to \order{\alpha^3 L_e^3}, and
the second-order corrections with the complete \order{\alpha^2 L_e}.
An extension of a program like \babayaga ~\cite{CarloniCalame:2000pz,CarloniCalame:2001ny,Balossini:2006wc}, which is currently exact at NLO with a matched QED shower,
to the level of NNLO for the hard process, while
keeping the correct soft-photon resummation,
would in principle provide the best comparison to the upgraded \bhlumi\ to establish the
technical precision of both programs at the $10^{-5}$
precision level%
\footnote{ The upgrade of the \bhlumi\ distributions will be
  relatively straightforward because its multi-photon 
  phase space is exact~\cite{Jadach:1999vf} for any number of photons.}.
However, a very good test of the technical precision of the upgraded \bhlumi\ would follow from the comparison between its results with EEX and CEEX matrix elements; for,
the basic multi-photon phase space integration module of \bhlumi\
was already well tested in Ref.~\cite{bhlumi-semi:1996}
and such a test can be repeated at an even higher-precision level. An alternative soluton, perhaps using the 
results in Refs.~\cite{frixione-2019,bertone-2019}, should also be pursued.
\par
\section{Synergies}
Historically, our exact ${\cal O}(\alpha^2 L)$ corrections were done for BHLUMI4 precision. They were then combined via crossing
with CEEX amplitude-based resummation in the MC KKMC for state-of-the-art 2f production in $e^+e^-$ annihilation. This realization of KKMC was then extended to KKMC-hh for Z production in $pp$ and $\bar{p}p$ collisions. This is now being extended to
MG5\_aMC@NLO/KKMC-hh~\cite{toappear} which would 
realize exact QCD NLO $\otimes$ exact ${\cal O}(\alpha^2 L)$ EW with 
matrix element matched parton showers. This chain of development illustrates the synergistic nature of progress in precision quantum field theory.
Indeed, when we add to BHLUMI the QED matrix element corrections of ${\cal O}\left(\alpha^2 L\right)$ and ${\cal O}\left(\alpha^3 L^3\right)$ we
already reduce $\delta N_\nu$ from $\delta{\cal L}/{\cal L}$ to 0.0015.\par
As a technical precision solution, we need to take CEEX to BHLUMI. Synergistically, for FCC-ee, this suggests taking CEEX to all
the EEX YFS realizations for LEP: for YFSWW3 \& KORALW~\cite{kandy-2001}, this is discussed in Refs.~\cite{MSkrzypek-2020,Jadach:2019a};
for YFSZZ~\cite{yfszz:1997} and BHWIDE, the work is in progress -- we do need sufficient theory resources. \par
The pay-off is a good one. For example, for $A_{FB}$ Ref.~\cite{jad-yost-afb} use CEEX via KKMC state-of-the-art 2f production
to achieve already the precision tag $\Delta\left[A_{FB}\right]_{IFI}\sim 10^{-4}$ on the IFI contribution to $A_{FB}$ which shows that 
the FCC-ee requirement is accessible. Similarly, as discussed in Ref.~\cite{MSkrzypek-2020}, the FCC-ee requirement of
a precision tag $\Delta M_W\sim 0.3 MeV$ in threshold and reconstruction impllies the CEEX extension of YFSWW \& KORALW is need in both cases. We note that, in Ref.~\cite{Jadach:2019a} this latter extension is in progress: the CEEX formalism is applied to
to $e^+e^- \rightarrow WW+ n\gamma \rightarrow 4f+ n'\gamma$. We note that contact between the results in Ref.~\cite{Jadach:2019a} and the usual 
Kleiss-Stirling spinor product based photon helicity infrared factors in CEEX proceeds via
\begin{equation}
ej^\mu_X(k_i)=eQ_X\theta_X\frac{2p^\mu_X}{2p_Xk_i} \rightarrow \sfac_{\sigma_i}\left(k_i\right) = eQ_X\theta_X\frac{b_{\sigma_i}\left(k_i,p_X\right)}{2p_Xk_i},
\end{equation}
with 
\begin{equation}
b_{\sigma}\left(k,p\right)=\sqrt{2}\frac{\bar{u}_\sigma\left(k\right)\not\!\!p u_\sigma\left(\zeta\right)}{\bar{u}_{-\sigma}\left(k\right)u_\sigma\left(\zeta\right)}, 
\end{equation}
where we follow the definitions in Refs.~\cite{Jadach:2019a,Jadach:2000ir}. The way forward is entirely open.\par
\section{Summary}
In summary,  we conclude that, with the appropriate resources, the path to $0.01\%$ precision for the FCC-ee luminosity (and the CLIC, ILC and CEPC luminosities) at the $Z$ peak is open
via an upgraded version of \bhlumi. It remains a matter for the appropriation of the required resources to support the attendant theoretical research -- a matter beyond our control. 
\par
\section*{Acknowledgments} This work was partially supported by the National Science Centre, Poland Grant No. 2019/34/E/\\
ST2/00457 and by The Citadel Foundation. We thank Patrick Janot for helpful discussion. Two of us (SJ and BFLW) thank Prof. Gian Giudice for the support and kind hospitality of the CERN TH Department. 
\setlength{\bibsep}{1.7pt}
\bibliography{Tauola_interface_design}{}
\bibliographystyle{utphys_spires}



\end{document}